# Freeform geometrical optics I: Principles

JUAN CAMILO VALENCIA-ESTRADA,[1*] AND JORGE GARCÍA-MÁRQUEZ[1]

[1] *Oledcomm SAS, Address 10-12, avenue de l'Europe, Vélizy, France, 78140*
*camilo.valencia@oledcomm.net*

**Abstract:** Lens design uses a calculation of the lens' surfaces that permit to obtain an image from a given object. A set of general rules and laws permits to calculate the essential points of the optical system such as distances, thickness, pupils, and focal distance among others. Now, the theory on which the classical lens design is based changes radically as our theoretical foundations do not rely on the classical ray tracing rules. We show that with the rules expressed in a reduced vector analytical solution set of equations, we can take into account all optical elements, i.e. refractive, reflective, catadioptric. These foundations permit to keep under control the system aberration budget in every surface. It reduces the computation time dramatically. The examples presented here were possible because of the versatility of this theoretical approach.

## 1. Introduction

Geometrical optic allows us to design optical systems by simply drawing a couple of rays having an object as the origin and an image as the final destination. These rays are called the chief ray and the marginal ray. The marginal ray departs from the most representative object point; it lies on the optical axis, travelling through the diaphragm edge. The chief ray departs from the outermost object point travelling through the diaphragm's center as shown in figure 1(a). Classical paraxial chief and marginal rays allow one to calculate the higher-order aberrations to control the image quality at large $F$-numbers. Thus, it is a starting point design for optimization of fast rotationally symmetric systems. Nonetheless, the classical paraxial theory has limitations when used to design systems with freeform surfaces. We should point out that all orders of spherical aberration cannot be controlled using the two-rays approach. However, in a high order approximation, all other classical types of aberrations can be examined. The higher order spherical aberration is mostly occurring in high numerical apertures or low $F$-number applications. In the case of wide-angle, large field-of-view applications, and refined chromatic corrections, the higher-order contributions of other aberrations are also relevant. Here we show a theory to design optical systems when compared to the classical geometrical optics one. This theory allows us maintain under control all the aberrations in freeform optical systems.

 The model does not have a fixed coordinate system. It depends on the optical paths and there is a single analytical solution that permits for the design of complex optical systems. Ray tracing can be used for assessment. The complete set of equations used in geometrical optics is replaced by a reduced set of vector formulas. This foundation can be easily explained and applied computationally.

 Geometrical optics is based in drawing rays from an arbitrary point in an object to the corresponding point in an image after crossing through the optical system. W. Snell and R. Descartes proposed the refraction and reflection laws in a time when vectors did not exist.

Analytic geometry is credited to Descartes and Fermat, thus geometrical optics just followed the mathematical knowledge available at that time. More than a century later, Gaussian optics approximation considered only points and rays in the close neighborhoods of the optical axis. In this approximation, the rays have small slopes when crossing the axis, which permits the sines of the angles to be replaced by the angles themselves [1-7]. This formulation is known as paraxial ray tracing and is valid when the angles are expressed in radians. Indeed, computation of small angles was simpler than trigonometric functions when computers were not yet invented.

A much more accurate method to design optical instruments consists in the successive application of the Snell-Descartes' law which requires an elementary geometry. This method gives estimation to the amount of aberrations introduced into the image by the optical system; examples of aberration's theories are Seidel's and Zernike's. E. Abbe introduced marginal rays, chief rays, and the stops. Both rays permit one to compute the image plane and the image height. Then vectors appeared. A. F. Möbius and W. R. Hamilton proposed them in a primitive form in the first half of the XIX century. Nevertheless, vector's rapid application was motivated by Maxwell's electromagnetic theory and its practical notation –due to J. W. Gibbs and O. Heaviside– permitted to reshape geometrical optics. That mathematical structure included two powerful mathematical tools: the vectors and the differential geometry as it is still used [2, 3].

This article is the first part of a series of articles dealing with optical design with freeform optics. The following two articles will treat on "Freeform geometrical optics II: From parametric representation to CAD/CAM" and "Freeform Geometrical Optics III: Optical System Design". Others dealing with optical assemblies, pupils and reference aperture's domain will hopefully follow.

## 2. Foundation principles

In image forming optical systems the light departing from an object $O$ in the object space is transferred to the image space where the image $I$ is formed. An example of an optical system depicting the bijective transformation correspondence having magnification is shown in Fig. 1(a). For the sake of simplicity, the current geometrical optics model is reduced to an object plane, a set of interfaces and an image plane. Here, we propose a method having the following two major differences when compared with current ones. The first difference is that its vector representation allows us to calculate analytically the optical system's surfaces. This is of great importance for getting a spherical aberration free system. The second, consists in its optimization method that can be briefly explained as a surface integral that assets the quality of the image. The most powerful result is that no rays are needed to get the best possible image.

### 2.1 Arbitrary optical path

Our method description requires a singlet transparent lens consisting of two or more refractive surfaces in a Cartesian space. For the description of the lens we will use the following notation: a) A sub-index 0 is attributed to the real-object subspace, b) the optics subspace corresponding to the lens is indicated by sub-index 1, and c) the real-image subspace is represented by sub-index 3. Let every subspace have a given refractive index $\{n_0, n_1, n_2\}$, where each index is a function of the wavelength, the temperature, or its position and direction in the space.

Let's consider a three-dimensional Cartesian space with coordinates $\{x, y, z\}$ where the lens is located. The Cartesian space is referenced to an absolute coordinates' origin $\{0, 0, 0\}$. This origin is the very optical system's observation point. Let's introduce an object $O$ in the space; we can assume that the object can be shifted along its reference optical path to different functional positions; i. e. the position could be expressed as a function of the reference aperture's domain –equivalent for the classical aperture stop– coordinates. The aperture's domain is defined as a reference closed bounded plane that define the optical system's parametric variables. The same is valid for the on-path image positions. For instance, this procedure can be used for designing astigmatic, cylindrical or progressive lenses. From all the points constituting an object, we select the most representative one that will be noted as $P_0$, having a position vector $\mathbf{p}_0 = [x_0, y_0, z_0]$. This point focuses our interest and can be seen in Fig. 1. We are going to create a reference optical path that will define the path a light ray traveling from the point $P_0$ to its image point $P_3$. In isotropic and homogeneous medias this is when the refractive indexes are constants. The optical path can be represented with optical vectors. We assume that a ray departs from $P_0$ to a point $P_1$ on the first lens surface $z_1(x_1, y_1)$ whose position vector is $\mathbf{p}_1 = [x_1, y_1, z_1]$. According to Snell-Descartes' law, the ray at $P_1$ is refracted and changes its direction to the second and unknown lens surface $z_2(x_2, y_2)$. On this second surface, the ray impinges an unknown point $P_2$ with an unknown position vector $\mathbf{p}_2 = [x_2, y_2, z_2]$. The ray is refracted again and directs to its destination at the known image point $P_3$ with position vector $\mathbf{p}_3 = [x_3, y_3, z_3]$. This arbitrary path is shown in Fig. 1. Position vector of points $P_0$ to $P_3$ in this arbitrary path are

$$\left.\begin{aligned}\mathbf{p}_0 &= [x_0, y_0, z_0]\\ \mathbf{p}_1 &= [x_1, y_1, z_1]\\ \mathbf{p}_2 &= [x_2, y_2, z_2]\\ \mathbf{p}_3 &= [x_3, y_3, z_3]\end{aligned}\right\}. \tag{1}$$

The vectors that describe the arbitrary segments are calculated by taking the position differences

$$\left.\begin{aligned}\mathbf{a}_0 &= \mathbf{p}_1 - \mathbf{p}_0\\ \mathbf{a}_1 &= \mathbf{p}_2 - \mathbf{p}_1\\ \mathbf{a}_2 &= \mathbf{p}_3 - \mathbf{p}_2\end{aligned}\right\}, \tag{2}$$

where its Euclidian distances -or norms- are $a_0 = \sqrt{\mathbf{a}_0 \bullet \mathbf{a}_0}$, $a_1 = \sqrt{\mathbf{a}_1 \bullet \mathbf{a}_1}$, and $a_2 = \sqrt{\mathbf{a}_2 \bullet \mathbf{a}_2}$.

Thus, the complete arbitrary optical path is obtained by adding the ray segment's distances multiplied by its respective refractive index: $\pm n_0 a_0 + n_1 a_1 \pm n_2 a_2$. It can be seen that in the case when either the object point or the image point are virtual, i.e. when they are not in their respective real subspaces, the optical trajectories are negative.

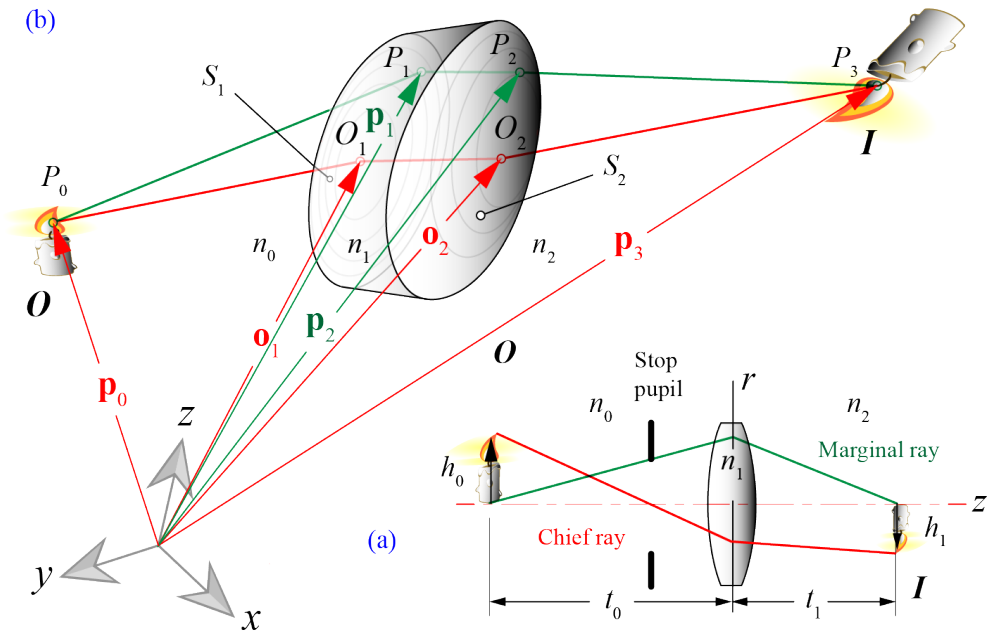

Fig. 1 Current and proposed optical variables in ray tracing. In (a) we represent the current way to describe all the variables that permits trace rays for image formation. Note that the chief and marginal rays describe the position of the inverted image and that the optical axis lies on a straight line that crosses the lens by its vertex. In (b) is the novel optical path tracing describing both, reference and arbitrary optical paths related to an arbitrary origin of coordinates. Note that the image should not be necessarily inverted as it could meet specific optical system's requirements. The optical axis does not lie on a straight line but are deliberately tilted to show the method's performance. All paths are vectored.

## 2.2 Reference optical path

The optical design process usually needs to answer the following questions: Where are placed the object, the image, and the optical system? What is the desired magnification? What is the maximum size of the optical system and its required resolution expressed in $F$-number? Finally, are there other constraints? According to the given answers, the designer determines if a freeform is needed. This method requires paraxial formulas to calculate the lens thickness, the object and the image distances. Once this step is done, the optical designer can proceed to calculate the vector positions of initial points $P_0$, $O_1$, $O_2$ y $P_3$.

We assume that there is a unique ray that departs from $P_0$ arrives to a known point $O_1$ on the first lens surface $S_1$ represented as the function $z_1(x_1, y_1)$ with known position vector $\mathbf{o}_1 = [x_{o1}, y_{o1}, z_{o1}]$. In said point the incident ray is refracted, continuing its trajectory to the second lens surface $S_2$ represented as the function $z_2(x_2, y_2)$ arriving at a known point $O_2$ with a known position vector $\mathbf{o}_2 = [x_{o2}, y_{o2}, z_{o2}]$. Again, the ray is refracted to continue its trajectory to the destination point $P_3$. The graphical description of this optical path is depicted in figure 1(a). From this, the reference vectors are

$$\left.\begin{array}{l}\mathbf{r}_0 = \mathbf{o}_1 - \mathbf{p}_0 \\ \mathbf{r}_1 = \mathbf{o}_2 - \mathbf{o}_1 \\ \mathbf{r}_2 = \mathbf{p}_3 - \mathbf{o}_2\end{array}\right\}. \qquad (3)$$

It is worth mentioning that in this method the reference optical path does not need to lie on the same straight line (i.e. the optical axis) or on the same plane as made in the two-dimensional or meridional ray-tracing methods. Its Euclidian norms are $r_0 = \sqrt{\mathbf{r}_0 \bullet \mathbf{r}_0}$, $r_1 = \sqrt{\mathbf{r}_1 \bullet \mathbf{r}_1}$, and $r_2 = \sqrt{\mathbf{r}_2 \bullet \mathbf{r}_2}$.

To guarantee that this proposed model satisfies Snell-Descartes' law in the first interface, the unit normal vector of the first surface in $O_1$ should satisfy the vector Snell-Descartes' condition [8]

$$\nabla z_1\big|_{O_1} = \tau\left(\pm n_1 \mathbf{r}_1 / \|\mathbf{r}_1\| \mp n_0 \mathbf{r}_0 / \|\mathbf{r}_0\|\right). \qquad (4)$$

Here $\nabla$ is the gradient operator evaluated at $O_1$, $\tau$ is a constant and $\|*\|$ is the norm of $*$. This condition states that a gradient vector is linearly dependent of the difference of optical unit vectors. The sign in Eq. (4) depends on the object position. The validity of Eq. (4) applies when $\nabla z_1 \neq \{0,0,0\}$. Otherwise, the anterior surface $S_1$ must be linearly transformed by using tilt, rotation or translation operators. This step can be automatically obtained for the first given surface. In other words: if the Snell-Descartes' vector condition in Eq. (4) is met, there is no need to calculate the partial derivatives in point $O_1$. However, if the condition is not met, the surface must be tilted to satisfy Eq. (4). This process involves the knowledge of partial derivatives for the normal at point $O_1$.

As tilt is now present in the system, $\mathbf{a}_0$ and $\mathbf{a}_1$ must be calculated again using Eq. (2) as well as their norms. More details on these transformations will be explained ahead. Thus, the Fermat's principle calculates the complete reference optical path just by adding the distance segments of the ray trajectories multiplied by its respective refraction index: $\pm n_0 r_0 + n_1 r_1 \pm n_2 r_2 = K$.

*2.3 Fermat's principle*

As a consequence of the Fermat's principle, it can be established that the time $t$ that every single ray takes to travel from an object point to an image point is always constant and equal to $T$. Fermat's principle is used to equal the arbitrary optical path with the reference optical path,

$$\pm n_0 a_0 + n_1 a_1 \pm n_2 a_2 = \pm n_0 r_0 + n_1 r_1 \pm n_2 r_2 = K. \qquad (5)$$

This approach is the key foundation of this theory as it abolishes the optical axis convention [6] in optical systems.

To automatically calculate the sign of each segment containing two signs in any refractive case, Eq. (5) can be expressed as

$$\text{Sign}(\mathbf{r}_0 \bullet \mathbf{r}_1)\, n_0\, a_0 + n_1\, a_1 + \text{Sign}(\mathbf{r}_1 \bullet \mathbf{r}_2)\, n_2\, a_2 = \text{Sign}(\mathbf{r}_0 \bullet \mathbf{r}_1)\, n_0\, r_0 + n_1\, r_1 + \text{Sign}(\mathbf{r}_1 \bullet \mathbf{r}_2)\, n_2\, r_2 = K,$$

(6)

This is a scalar equation. In the following section we are going to transform it into a vector equation by using the vector form of Snell-Descartes' law.

*2.4 Vector form of Snell-Descartes' law*

By using the vector form of Snell-Descartes' law, it is possible to find a unit vector $\mathbf{v}_1$ in the direction of a refracted ray. Again, by using Snell-Descartes' vector law we can calculate the unit vector in the direction of the refracted ray once both the unit vector $\mathbf{n}_1$ normal to the optical interface and the unit incident vector $\mathbf{v}_0$ are known. Thus,

$$\mathbf{v}_0 = \text{Sign}(\mathbf{r}_0 \bullet \mathbf{r}_1)\, \frac{\mathbf{a}_0}{\sqrt{\mathbf{a}_0 \bullet \mathbf{a}_0}},$$

(7)

This formula permits one to switch unit vectors direction when the object point is virtual. The unit vector $\mathbf{n}_1$ normal to the first interface is defined as

$$\mathbf{n}_1 = -\frac{\left.\dfrac{\partial \mathbf{p}_1}{\partial x} \times \dfrac{\partial \mathbf{p}_1}{\partial y}\right.}{\left\|\dfrac{\partial \mathbf{p}_1}{\partial x} \times \dfrac{\partial \mathbf{p}_1}{\partial y}\right\|_{x\,y}}.$$

(8)

By applying the vector form of Snell-Descartes' law using dot products in the arbitrary point $P_1$

$$\mathbf{v}_1 = \frac{n_0}{n_1}\left(\mathbf{v}_0 - (\mathbf{n}_1 \bullet \mathbf{v}_0)\,\mathbf{n}_1\right) - \left(\sqrt{1 - \frac{n_0^2}{n_1^2}\left(1 - (\mathbf{n}_1 \bullet \mathbf{v}_0)^2\right)}\right)\mathbf{n}_1 = \begin{bmatrix} a \\ b \\ c \end{bmatrix},$$

(9)

thus, the directional cosines $a$, $b$, and $c$ of said unit vector can be obtained. This unit vector has the exact same direction of the ray propagating in the interior lens (internal ray). It is important to keep in mind the identity $a^2 + b^2 + c^2 = 1$. Now, it is possible to determine the equation of the line that contains the internal ray $(x_2 - x_1)/a = (y_2 - y_1)/b = (z_2 - z_1)/c$, the internal arbitrary vector can be expressed by $\mathbf{a}_1 = a_1\, \mathbf{v}_1$.

## 3. The basic equation system and its solution

When for any point at the front surface, the first surface and its position vector $P_1$ are known, then the second surface is found. For doing this, the Fermat principle is applied and spherical aberration is removed for the conjugate points $P_0$ and $P_3$. Up to now, our model consists in a three-equations and three unknowns system: a) the vector form of Snell-Descartes' equation (9), b) a vector equation $\mathbf{a}_1 = a_1\, \mathbf{v}_1$ that represents the internal arbitrary ray and, c) the Fermat principle in scalar form, Eq. (6). With these equations we can calculate the incident point $P_2$ of the internal ray's position vector in such a way that once refracted by the second lens interface the ray arrives exactly at the image point $P_3$. Next, we explain how the solution can be found. We convert the scalar equation (6) in a vector one by multiplying it by $\mathbf{v}_1$. Its result has three scalar equations, one for each coordinate. By regrouping terms, we obtain

$$\text{Sign}(\mathbf{r}_1 \bullet \mathbf{r}_2)\, n_2\, a_2 \mathbf{v}_1$$
$$= \text{Sign}(\mathbf{r}_0 \bullet \mathbf{r}_1)\, n_0\, r_0 \mathbf{v}_1 + n_1 r_1 \mathbf{v}_1 + \text{Sign}(\mathbf{r}_1 \bullet \mathbf{r}_2)\, n_2 r_2 \mathbf{v}_1 - \text{Sign}(\mathbf{r}_0 \bullet \mathbf{r}_1)\, n_0 a_0 \mathbf{v}_1 - n_1 a_1 \mathbf{v}_1. \tag{10}$$

Then, by dividing both sides in Eq. (10) by the refractive index $n_2$ and substituting the last term $a_1 \mathbf{v}_1$ by $\mathbf{a}_1$ we can simplify it as,

$$((\mathbf{p}_3 - \mathbf{p}_2) \bullet (\mathbf{p}_3 - \mathbf{p}_2))\, \mathbf{v}_1 = \left( k_1 \mathbf{v}_1 - \frac{n_1}{n_2}(\mathbf{p}_2 - \mathbf{p}_1) \right)^2. \tag{11}$$

where the scalar is $k_1 = \text{Sign}(\mathbf{r}_0 \bullet \mathbf{r}_1)(n_0 / n_2)(r_0 - a_0) + (n_1 / n_2)r_1 + \text{Sign}(\mathbf{r}_1 \bullet \mathbf{r}_2)r_2$. This equation can be rearranged in such a way that its left side contains only the unknown dot product terms as they depend on the unknown position vector $\mathbf{p}_2$.

$$(\mathbf{p}_2 \bullet \mathbf{p}_2)\, \mathbf{v}_1^2 - 2(\mathbf{p}_2 \bullet \mathbf{p}_3)\, \mathbf{v}_1^2 = \left( k_1 \mathbf{v}_1 - \frac{n_1}{n_2}(\mathbf{p}_2 - \mathbf{p}_1) \right)^2 - (\mathbf{p}_3 \bullet \mathbf{p}_3)\, \mathbf{v}_1^2. \tag{12}$$

These two dot product terms can be transformed into vector functions not containing the dot product.

$$(\mathbf{p}_2 \bullet \mathbf{p}_2)\, \mathbf{v}_1^2 = \mathbf{p}_2^2 - 2\mathbf{p}_2(\mathbf{p}_1 - (\mathbf{p}_1 \bullet \mathbf{v}_1)\mathbf{v}_1) + \mathbf{p}_1^2 + (\mathbf{p}_1 \bullet \mathbf{p}_1)\mathbf{v}_1^2 - 2(\mathbf{p}_1 \bullet \mathbf{v}_1)\mathbf{p}_1 \mathbf{v}_1. \tag{13}$$

Next, we present the second step to transform the second left side term in Eq. (12).

$$-2(\mathbf{p}_2 \bullet \mathbf{p}_3)\, \mathbf{v}_1^2 = -2\big((\mathbf{p}_3 \bullet \mathbf{v}_1)(\mathbf{p}_2 - \mathbf{p}_1)\mathbf{v}_1 + (\mathbf{p}_1 \bullet \mathbf{p}_3)\mathbf{v}_1^2\big). \tag{14}$$

Substituting summands of Eqs. (13) and (14) in the left side of Eq. (12) and by simplifying, we obtain

$$\left( \frac{n_2^2 - n_1^2}{n_2} \right)\mathbf{a}_1^2 - 2k_3 \mathbf{v}_1 \mathbf{a}_1 + n_2(k_2 - k_1^2)\mathbf{v}_1^2 = 0, \tag{15}$$

with the scalars $k_2 = (\mathbf{p}_3 - \mathbf{p}_1) \bullet (\mathbf{p}_3 - \mathbf{p}_1)$ and $k_3 = n_2(\mathbf{v}_1 \bullet (\mathbf{p}_3 - \mathbf{p}_1)) - n_1 k_1$. It is now possible to find the solution for the unknown internal unit vector $\mathbf{a}_1$. It is convenient to use the next expression that is completely in a vector form

$$\mathbf{a}_1 = \frac{n_2}{n_2^2 - n_1^2}\left( k_3 + \text{Sign}(\mathbf{r}_1 \bullet \mathbf{r}_2)\sqrt{k_3^2 - (n_2^2 - n_1^2)(k_2 - k_1^2)} \right)\mathbf{v}_1. \tag{16}$$

Note that the sign does not switch and that this solution is not determined when the denominator is null. The sign of the square root depends on the position of the image point.

With the result in Eq. (16) it is possible to find the unknown position vector that let us write parametrically the geometry of the second surface in such a way that the point image $P_3$ is aberrations-free.

$$\mathbf{p}_2 = \mathbf{p}_1 + \mathbf{a}_1 = \mathbf{p}_1 + \frac{n_2}{n_2^2 - n_1^2}\left(k_3 + \text{Sign}(\mathbf{r}_1 \bullet \mathbf{r}_2)\sqrt{\left(k_3^2 - (n_2^2 - n_1^2)(k_2 - k_1^2)\right)}\right)\mathbf{v}_1. \quad (17)$$

Equation (17) is valid for any object point $P_0$ no matter whether is real or virtual, the same that for any real or virtual image point $P_3$ as well as for any combination of conjugated points, with the condition

$$-\frac{\partial \mathbf{p}_2}{\partial x} \times \frac{\partial \mathbf{p}_2}{\partial y}\bigg|_{\substack{x\\y}} \neq [0,0,0] \ \forall (x,y) \subset \text{Aperture}, \quad (18)$$

that guarantees avoiding discontinuities and auto-intersections. It is also valid in the case of lenticular or discontinuous anterior surfaces (i.e. lenticular lenses, Fresnel lenses, etc). In the case when the first surface $S_1$ is discontinuous or piecewise, the proposed method works out for finding the second discontinuous surface $S_2$, by calculating the borders of each continuous sector. The condition given in Eq. (18) has to be fulfilled for the aperture of each sector.

## 4. Different optic cases

We consider here that Eq. (17) needs different signs depending on refraction or reflection. These changes are due to changes in sign's rules. For instance, according to the classical Snell formulation, if $n_2 = -n_1$, but with this new formulation it could be interpreted as a refraction in a meta-surface. Direction of reflected ray or refracted ray is the same but in opposite sense.

A general rationalized solution of Eq. (17) is used for a system with a back refractive interface:

$$\mathbf{p}_2 = \mathbf{p}_1 + \mathbf{a}_1 = \mathbf{p}_1 + \left(\frac{n_2 G}{V - s_3\sqrt{V^2 - (n_2^2 - n_1^2)G}}\right)\mathbf{v}_1. \quad (19)$$

With this equation division by zero is avoided in Eq. (17). The sign $s_3$ in the denominator depends on the image position, direction and magnification. When we want to design an optical system with a back reflective freeform interface, then Eq. (19) reduces to

$$\mathbf{p}_2 = \mathbf{p}_1 + \mathbf{a}_1 = \mathbf{p}_1 + \left(\frac{n_2 G}{2V}\right)\mathbf{v}_1. \quad (20)$$

The internal unit vector $\mathbf{v}_1$ is calculated with $\mathbf{v}_0 = s_4\, \mathbf{a}_0 / \sqrt{\mathbf{a}_0 \bullet \mathbf{a}_0}$. The sign $s_4$ depends on the object's position and direction. Note that a dioptric surface with a virtual object corresponds to a catoptric surface with a real object.

The recurrent variables can be represented by

$$A = k_1 = s_0 (n_0 / n_2)(r_0 - a_0) + (n_1 / n_2) r_1 + s_2 r_2,$$
$$G = k_2 - k_1^2 = (\mathbf{p}_3 - \mathbf{p}_1) \bullet (\mathbf{p}_3 - \mathbf{p}_1) - A^2, \qquad (21)$$
$$V = k_3 = n_2 (\mathbf{v}_1 \bullet (\mathbf{p}_3 - \mathbf{p}_1)) - n_1 A.$$

The sign $s_0$ depends on the object position and direction. The sign $s_2$ in $A$ depends on the image position, direction and magnification. This solution is still valid even if auto-intersections or loss of continuity of the solved surface appear, provided these auto-intersections or discontinuous sectors are out of the aperture's covering sector. This can be easily assessed by the condition given in Eq. (18) or by ray drawing or tracing. The set of previous equations is still valid in the case the rays cross internally, as shown in Fig. (2). When the internal fan of rays is divided in two different fans –from which one presents rays' inversion– the diaphragm can be replaced in order to block out the undesired fan. As a result, auto intersections and discontinuities are avoided.

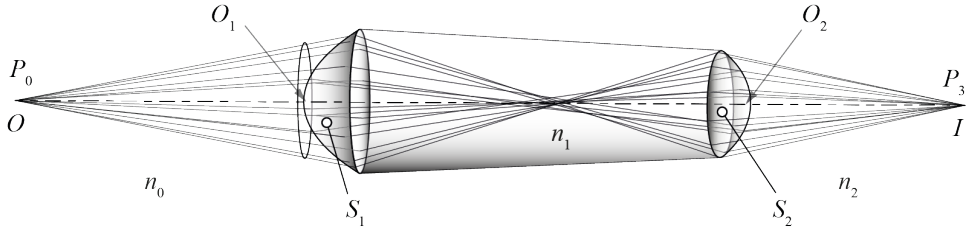

Fig. 2 A lens presenting an internal fan of rays crossing internally. The vector formulas for freeform optical systems design are useful to design on axis systems with a collinear reference optical path.

According to the rule of signs, design of two-interface optical systems is classified as follows:

a) Dioptric case when both surfaces are refractive. Then the dioptric solution in Eq. (19) is used to design a pure refractive system (*i. e.* a lens).

b) A catadioptric case when a first surface is refractive and a second surface is reflective, in such a case the catoptric solution in Eq. (20) is used.

c) A catadioptric case in which a first surface is reflective and a second surface refractive. Then, the dioptric solution described by Eq. (19) applies. Finally,

d) A Catoptric case when both surfaces are reflective. In this last case Eq. (20) is used to design a 2-mirrors system.

We conclude that the vector general solution expressed in Eqs. (19-21) allows one to design a two-interface optical system consisting in refractive, reflective or any combination of these surfaces.

Equations (19-21) are valid for any object point $P_0$ no matter whether is real or virtual. The same is valid for any real or virtual image point $P_3$ as well as for any combination of conjugated points.

## 5. Reversibility principle

Let's suppose a lens that forms a perfect image of an object point, the lens is said to be reversible. This means that the image point $I$ becomes a new object point $\hat{O}$, and that the object point $O$ becomes a new image point $\hat{I}$. This principle is of great importance as it will be shown in a forthcoming part of this series on Freeform Geometrical Optics. The solution to this problem is simplified in the case the second surface is given and then by calculating the position vector of the first surface as described ahead.

Similarly, as for the second interface, the unit normal vector of the second surface in $O_2$ should satisfy the vector condition $\nabla z_2|_{O_2} = \tau \left( \pm n_2 \mathbf{r}_2 / \|\mathbf{r}_2\| \mp n_1 \mathbf{r}_1 / \|\mathbf{r}_1\| \right)$.

As said before, this condition states that a gradient vector is linearly dependent of the difference of optical unit vectors. Otherwise, the posterior surface $z_2(x_2, y_2)$ must be linearly transformed by using tilt, rotation, or translation operators. This step can be automatically obtained for the given back surface.

Snell-Descartes' vector law allows you to calculate the unit vector $\mathbf{v}_1$ in the direction of the incident ray once both the normal unit vector $\mathbf{n}_2$ to the optical interface and the unit refracted vector $\mathbf{v}_2$ are known. Thus, $\mathbf{v}_2 = \text{Sign}(\mathbf{r}_1 \bullet \mathbf{r}_2) \mathbf{a}_2 / \sqrt{\mathbf{a}_2 \bullet \mathbf{a}_2}$. This formula permits you to switch unit vectors direction when the image point is virtual. The unit vector $\mathbf{n}_2$ normal to the second interface is

$$\mathbf{n}_2 = -\frac{\dfrac{\partial \mathbf{p}_2}{\partial x} \times \dfrac{\partial \mathbf{p}_2}{\partial y}}{\left\| \dfrac{\partial \mathbf{p}_2}{\partial x} \times \dfrac{\partial \mathbf{p}_2}{\partial y} \right\|_{x}^{y}}. \tag{22}$$

By applying the vector form of Snell-Descartes' law in the arbitrary point $P_2$

$$\mathbf{v}_1 = \frac{n_2}{n_1} \left( \mathbf{v}_2 - (\mathbf{n}_2 \bullet \mathbf{v}_2) \mathbf{n}_2 \right) - \left( \sqrt{1 - \frac{n_2^2}{n_1^2} \left( 1 - (\mathbf{n}_2 \bullet \mathbf{v}_2)^2 \right)} \right) \mathbf{n}_2, \tag{23}$$

The solution for $\mathbf{p}_1$ in a refractive case can be obtained by using the equation (19) taking care in using the signs defined for the refractive case. We apply the following variable change: (a) direction switch of the inner vector $\mathbf{v}_1$; (b) by swapping vectors $\mathbf{p}_1$ by $\mathbf{p}_2$, and $\mathbf{p}_3$ by $\mathbf{p}_0$; (c) by swapping vectors $\mathbf{r}_1$ by $\mathbf{r}_0$, and refractive indices $n_2$ by $n_0$, to obtain

$$\mathbf{p}_1 = \mathbf{p}_2 - \frac{n_0 G}{V - \text{Sign}(\mathbf{r}_1 \bullet \mathbf{r}_0)\sqrt{V^2 - (n_0^2 - n_1^2)G}} \mathbf{v}_1, \quad (24)$$

The recurrent variables can be represented by

$$\begin{aligned} A &= \text{Sign}(\mathbf{r}_2 \bullet \mathbf{r}_1)(n_2/n_0)(r_2 - a_2) + (n_1/n_0)r_1 + \text{Sign}(\mathbf{r}_1 \bullet \mathbf{r}_0)r_0, \\ V &= n_0(-\mathbf{v}_1 \bullet (\mathbf{p}_0 - \mathbf{p}_2)) - n_1 A, \\ G &= (\mathbf{p}_0 - \mathbf{p}_2) \bullet (\mathbf{p}_0 - \mathbf{p}_2) - A^2. \end{aligned} \quad (25)$$

Equation (24) is valid for any object point $P_3$ no matter whether it is real or virtual, the same that for any real or virtual image point $P_0$ as well as for any combination of conjugated points. There is one condition: the existence of a plane tangent to all the aperture points that guarantee against discontinuities or auto-intersections:

$$\left. -\frac{\partial \mathbf{p}_1}{\partial x} \times \frac{\partial \mathbf{p}_1}{\partial y} \right|_{x,y} \neq [0,0,0] \quad \forall (x,y) \subset \text{Aperture}, \quad (26)$$

that guarantees avoiding discontinuities and auto-intersections. It is also valid in the case of lenticular or discontinuous anterior surfaces (i. e lenticular lenses, Fresnel lenses, etc). In the case when the first surface $S_2$ is discontinuous or piecewise, the proposed method works out for finding the second discontinuous surface $S_1$, by calculating the borders of each continuous sector. The condition given in Eq. (26) has to be fulfilled for the aperture of each sector.

## 6. Surface optimization to obtain the best image

The description of this section needs a single two-surface lens. The reader will realize that this description is easily generalized to other kinds of optical systems. We assume that this lens is designed by following our method exposed here, i.e. that the lens is defined in a three-dimensional vector optical space. Note that the wave front has associated a propagation vector. For this description we require two object points $O$ and $O'$ and their corresponding image points $I$ and $I'$. As we have stated before, the object is not necessarily planar, in fact, eventually it could be planar, the same as its image; it is for this reason that the authors do not employ the image height (refer to figure 3). From the previous section, we saw that any object point $O$ has a perfect image $I$. Nevertheless, for all the other object points, their respective image could not be perfect. Now, consider a spherical wave front propagating from the object point $O'$ going through the optical system to finally arrive to its corresponding image point $I'$. We suppose that there is one optical path that arrives to $I'$. This is our new reference optical path. This optical path can be seen as a ray that departing from $O'$ arrives to the first surface at a point $P_1'$ with a position vector $\mathbf{p}_1'$. This ray is refracted by the interface $S_1$ according to the Snell-Descartes' law to point the second surface $S_2$. When this ray impinges the second surface at the point $P_2'$ with position vector $\mathbf{p}'_2$, it is refracted again to point the ideal image point $I'$ with position vector $\mathbf{p}'_4$ as seen in Fig. 3. We need to calculate the coordinates of the position vector $\mathbf{p}'_2$, once this coordinates are known we use the simplest vector representation of Snell-Descartes' law that permits us to calculate the normal at a point if the incident and refracted optical vectors are known. This normal vector $\mathbf{n}'_2$ in the point $P'_2$ is $\mathbf{n}'_2 = -(\pm n_2\,\mathbf{v}'_1 \mp n_1\,\mathbf{v}'_1)/\sqrt{(\pm n_2\,\mathbf{v}'_1 \mp n_1\,\mathbf{v}'_1)\bullet(\pm n_2\,\mathbf{v}'_1 \mp n_1\,\mathbf{v}'_1)}$. Nevertheless, it is also possible to calculate the normal by determining the surface $S_2'$ that contains the point $P'_2$. This surface corrects all aberration introduced by the first surface $S_1$ in such a way that the point $I'$ is a perfect image. Equation (18) is evaluated with the parameters found for the new reference optical path. Let's suppose that $z'_2$ is required, then its normal at $P'_2$ is alternately calculated by $\mathbf{n}'_2 = -\left[\partial \mathbf{p}'_2/\partial x \times \partial \mathbf{p}'_2/\partial x\right]/\sqrt{(\partial \mathbf{p}'_2/\partial x \times \partial \mathbf{p}'_2/\partial x_1)\bullet(\partial \mathbf{p}'_2/\partial x \times \partial \mathbf{p}'_2/\partial x)}$. Finally, it is worth noting that when

$$\iint_{stop} \|\mathbf{n}_2 \times \mathbf{n}'_2\| ds \to 0 \quad \text{or} \quad \iint_{stop} ((\mathbf{n}_2 \times \mathbf{n}'_2)\bullet(\mathbf{n}_2 \times \mathbf{n}'_2)) ds \to 0, \tag{27}$$

the image of the points constituting the object is perfect. Without loss of generality this closed surface integrals can be equivalently described as

$$\text{Max}\left\{\sum_i^s |\mathbf{n}_{2_i}\bullet\mathbf{n}'_{2_i}|\right\} \quad \text{or} \quad \text{Min}\left\{\sum_i^s ((\mathbf{n}_{2_i} \times \mathbf{n}'_{2_i})\bullet(\mathbf{n}_{2_i} \times \mathbf{n}'_{2_i}))\right\}, \tag{28}$$

where the sum takes $s$ samples and $i$ is the counting index.

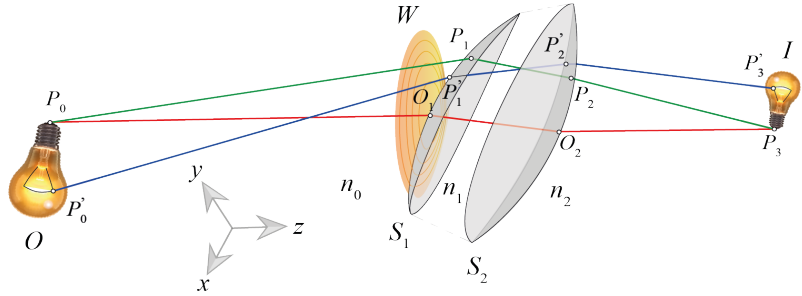

Fig. 3. Two different object's positions are depicted here: $P_0$ and $P'_0$ with their respective images $P_3$ and $P'_3$. We can see in blue the external reference optical paths following points $P'_0$ to $P'_3$. The external reference paths allow us for indirectly measuring aberrations in the image by calculating the ideal normal $\mathbf{n}'_2$ in $P'_2$. On the opposite side, only one main reference optical path (here, in red) is needed to design the lens. The lens has two convex surfaces $S_1$ and $S_2$. The wave front $W$ entering the first surface is also depicted. Note that both surfaces are not parallel but tilted and that its thickness is larger at bottom in turn introducing a prism. The refractive index between surfaces $S_1$ and $S_2$ is $n_1$ and $n_2 = n_0$. The versatility of this designing method permits the introduction of tilt or change of the distribution of the image.

In figure 4, we show an example of an off-axis lens as the vertex $O_1$ of the revolution paraboloid $S_1$ is not in the aperture lens. As we have already explained, this method allows high flexibility when designing optical systems. In this example, we have arbitrarily decided to place the point $O_1$ on the coordinate's origin $\{0, 0, 0\}$. The whole system assembly must have an absolute coordinate's origin placed anywhere in the space, as specified in any drawing. Optical design software (for instance Oslo®) usually considers a representative object's point as the absolute origin point. However, in the model shown here, the election of the coordinate's origin is completely flexible.

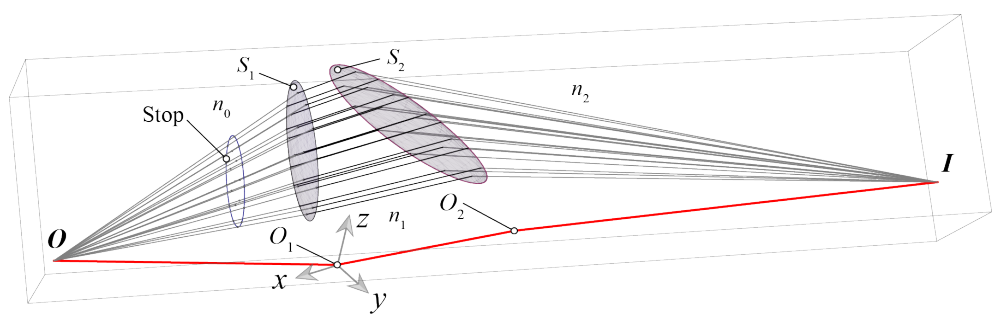

Fig. 4. An example of a freeform lens designed with an external reference path. The off-axis lens has an off-axis reference aperture's domain, the lens' surfaces $S_1$ and $S_2$ are highly tilted between them. In this example $S_1$ is an off-axis section of a revolution paraboloid. The aperture's domain rim defines the surface's edges. The reference optical path that consists of three segments is in red.

In figure 5, we show an example of a catadioptric system in which the object $O$ is virtual. The aperture's domain limits the entrance of rays into the system. Note that the aperture's domain is in contact with the vertex of the refractive front and tilted surface. The second surface is reflective and convex. It reflects the incoming rays into the image point $I$.

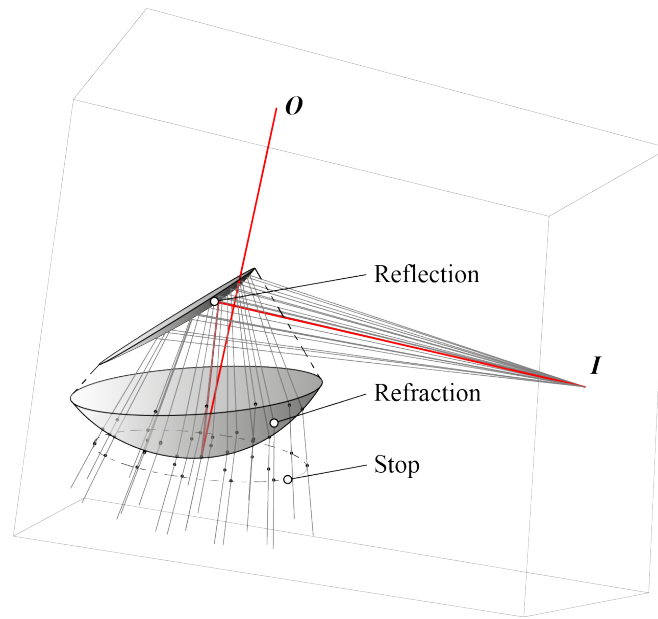

Fig. 5. A more complex example of catadioptric-design with the proposed method in which the object ***O*** is virtual. The aperture limits the entrance of converging rays into the optical system. The second surface is not refractive but reflective and deviates the rays into the image ***I***. The reference optical path is indicated in red.

## 7. Method's error estimation

The parametric surface's expressions obtained by using our method have hundreds of terms and are not suited for most applications. Nevertheless, we can obtain a cloud of thousands of surface's points (say $S_2$) that can be fitted to an explicit function such as depicted in figure 6(a). The representation error is the difference between the parametric and the explicit representation surfaces. We show this in figure 6(b). The brownish-red surface is the explicit representation surface; it overlaps the parametric surface (in green). (c) The quantitative error of this difference is depicted by the cloud of points.

By increasing the terms of the explicit function, we can reduce the approximation error. By increasing the terms of the explicit function –from which the two-dimensional Taylor, the cylindrical, or the Zernike polynomials are examples– we can reduce the approximation error. By calculating the wavefronts before and after each interface, these representations permit one to obtain the aberration contribution for each surface in the optical system. Some authors recommend representing freeform surfaces by using a conic base plus Zernike's deformation polynomials [9-11]. A clear correspondence exists between deformation polynomials and its respective contribution of aberrations only if it is near the pupil, due to sub-apertures and restricted ray bundle cross sections.

Once the explicit continuous (or piecewise) function is obtained, the surface can be exported to any Computer Assisted Design (CAD) or Computer Assisted Manufacture (CAM) software ready to manufacture.

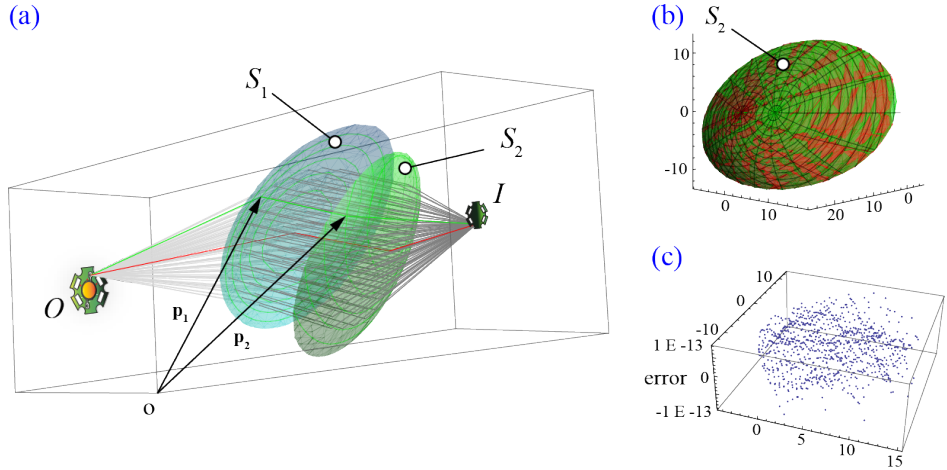

Fig. 6. Freeform optical system. (a) A single prismatic and refractive two-surfaces optical system images an LED. Its $S_2$ surface was calculated by using Eq. (18). (b) This surface $S_2$ is shown in green (parametric one) and in red, its best fitting (explicit one) is shown. The explicit surface was fitted with Zernike's polynomials. (c) After merging both surfaces, its difference is plotted in a cloud of uniformly distributed sample points. Note the insignificance of the residual error (expressed in mm).

## 8. Final discussions

Geometrical optics as it was conceived four hundred years ago is intuitive and was adapted to hand calculations. This classical geometrical optics theory evolved mainly in the 19[th] and 20[th] century in such a way it is possible to compute complicated rays tracing. Notwithstanding, a question can be asked: How can we obtain the best optical system with the minimal number of surfaces and with aberrations reduced to the diffraction limit or near this limit? Here we have explained how to obtain such a system with a vector method whose basis are the Snell-Descartes' law and the Fermat principle. The advantages of our method are: a) the optical system can be located in any place in space; as a consequence, the optical axis is not required. The designer can obtain any arbitrary topology sought after. b) Even if no ray tracing is required as the optical design is based on optical paths, the ray tracing for visual purposes can be displayed. c) No cardinal points are required. d) The analytical solution -that consists of a vector formula, Eq. (18) covers all kind of optical systems: dioptric, catoptric, and catadioptric systems. e) Every time a new surface is designed all the aberrations are automatically reduced in parallel by following the principle expressed in Eq. (27) or in Eq. (28). f) The last two points implies a dramatic reduction of computing time. g) The edge of each surface is a parametrical function of the aperture's domain's edge parameters. Finally, chromatic aberrations are corrected by combining different optical materials.

In [12], Miñano et al demonstrated the feasibility of the well establish Simultaneous Multiple Design (SMS) non-image forming optical design method [13] in image forming applications. Analytic vector solution that uses an optimization method based in Eq. (28) or alternatively Eq. (27), can also be used in both, non-image and image forming systems. The main difference between SMS method, as shown in [12, 13] and the method here presented is found on the optimization method. SMS compares local angles against ideal angles for all surface sampled points while the vector method compares local normals against ideal normals. In other words, SMS method is based in the ideal wave fronts while the vector method is based in ideal interfaces. Equation (28) is highly relevant as it permits to minimize the lens' aberrations. As the integral in (27) tends to zero, aberrations will reduce dramatically. The method here proposed has been thought as a new tool for optical design.

Nonetheless, the obtained parametric surface(s) have an extended mathematical representation that it makes difficult to express the result. Fortunately, this inconvenience can be easily resolved by transforming the parametric expression into an explicit one. Manufacturability of this kind of surfaces will appear in the second part of Freeform Geometrical Optics series, while generalization of this vector method to several surfaces systems is coming in the third part of this series. Other parts dealing with testing, stops and the newly introduced reference aperture's domain are under preparation.

## Disclosures

## Acknowledgments

This project has received funding from the European Union's Horizon 2020 research and innovation program under grant agreement No. 761329. The theoretical model explained here is mainly the result of the authors' independent work motivated by WORTECS project specifications. Both authors are in debt to Alexander Benson for the final English review of this manuscript. Authors are in debt to Dominic O'Brien and Grahame Faulkner (University of Oxford) for some fruitful discussions on CPC's. They are also in debt to Olivier Boucher (Orange Labs) for discussions on optical radiation requirements.

**Annex**

Here, we present the deductive steps to obtain the vector identities (equations 13 and 14)

$$(\mathbf{p}_2 \bullet \mathbf{p}_2) \mathbf{v}_1^2 = \mathbf{p}_2^2 - 2\mathbf{p}_2 (\mathbf{p}_1 - (\mathbf{p}_1 \bullet \mathbf{v}_1)\mathbf{v}_1) + \mathbf{p}_1^2 + (\mathbf{p}_1 \bullet \mathbf{p}_1)\mathbf{v}_1^2 - 2(\mathbf{p}_1 \bullet \mathbf{v}_1)\mathbf{p}_1 \mathbf{v}_1. \qquad (13)$$

$$-2(\mathbf{p}_2 \bullet \mathbf{p}_3) \mathbf{v}_1^2 = -2\left((\mathbf{p}_3 \bullet \mathbf{v}_1)(\mathbf{p}_2 - \mathbf{p}_1)\mathbf{v}_1 + (\mathbf{p}_1 \bullet \mathbf{p}_3)\mathbf{v}_1^2\right). \qquad (14)$$

The two dot product terms in the left side of each equation can be transformed into vector functions not containing the dot product. For the first term, vectors $\mathbf{p}_2$ and $\mathbf{v}_1$ are evaluated according to Eqs. (1) and (9) respectively, obtaining

$$(\mathbf{p}_2 \bullet \mathbf{p}_2) \mathbf{v}_1^2 = (x_2^2 + y_2^2 + z_2^2) \begin{bmatrix} a^2 \\ b^2 \\ c^2 \end{bmatrix} = \begin{bmatrix} x_2^2 a^2 + y_2^2 a^2 + z_2^2 a^2 \\ x_2^2 b^2 + y_2^2 b^2 + z_2^2 b^2 \\ x_2^2 c^2 + y_2^2 c^2 + z_2^2 c^2 \end{bmatrix}. \qquad (A1)$$

Position vector coordinates $\mathbf{p}_2$ that does not have a place in the vector in Eq. (A1) must be substituted. In this manner the variable $y_2$ and $z_2$ in the first raw, the variables in the second raw $x_2$ and $z_2$ and finally the variables in the third raw $x_2$ and $y_2$ must be changed to $x_2$, $y_2$, and $z_2$ respectively. To do this, we change variables required in each raw by using $(x_2 - x_1)/a = (y_2 - y_1)/b = (z_2 - z_1)/c$.

$$(\mathbf{p}_2 \bullet \mathbf{p}_2)\,\mathbf{v}_1^2 = \begin{bmatrix} a^2 x_2^2 + a^2 \left(\dfrac{b\,(x_2 - x_1)}{a} + y_1\right)^2 + a^2 \left(\dfrac{c\,(x_2 - x_1)}{a} + z_1\right)^2 \\ b^2 \left(\dfrac{a\,(y_2 - y_1)}{b} + x_1\right)^2 + b^2 y_2^2 + b^2 \left(\dfrac{c\,(y_2 - y_1)}{b} + z_1\right)^2 \\ c^2 \left(\dfrac{a\,(z_2 - z_1)}{c} + x_1\right)^2 + c^2 \left(\dfrac{b\,(z_2 - z_1)}{c} + y_1\right)^2 + c^2 z_2^2 \end{bmatrix},$$

(A2)

$$= \begin{bmatrix} a^2 x_2^2 + \left(b\,(x_2 - x_1) + a\,y_1\right)^2 + \left(c\,(x_2 - x_1) + a\,z_1\right)^2 \\ \left(a\,(y_2 - y_1) + b\,x_1\right)^2 + b^2 y_2^2 + \left(c\,(y_2 - y_1) + b\,z_1\right)^2 \\ \left(a\,(z_2 - z_1) + c\,x_1\right)^2 + \left(b\,(z_2 - z_1) + c\,y_1\right)^2 + c^2 z_2^2 \end{bmatrix}.$$

Expanding the right side in Eq. (A2) and using the identity $a^2 + b^2 + c^2 = 1$, Eq. (A2) is algebraically reduced to

$$(\mathbf{p}_2 \bullet \mathbf{p}_2)\,\mathbf{v}_1^2 = \mathbf{p}_2^2 + 2\mathbf{p}_2 \begin{bmatrix} a(\mathbf{p}_1 \bullet \mathbf{v}_1) - x_1 \\ b(\mathbf{p}_1 \bullet \mathbf{v}_1) - y_1 \\ c(\mathbf{p}_1 \bullet \mathbf{v}_1) - z_1 \end{bmatrix} + \mathbf{p}_1^2 + (x_1^2 + y_1^2 + z_1^2)\begin{bmatrix} a^2 \\ b^2 \\ c^2 \end{bmatrix} - 2\mathbf{v}_1 \begin{bmatrix} a x_1^2 + b x_1 y_1 + c x_1 z_1 \\ b y_1^2 + a x_1 y_1 + c y_1 z_1 \\ c z_1^2 + a x_1 z_1 + b y_1 z_1 \end{bmatrix},$$

$$= \mathbf{p}_2^2 - 2\mathbf{p}_2\left(\mathbf{p}_1 - (\mathbf{p}_1 \bullet \mathbf{v}_1)\mathbf{v}_1\right) + \mathbf{p}_1^2 + (\mathbf{p}_1 \bullet \mathbf{p}_1)\mathbf{v}_1^2 - 2(\mathbf{p}_1 \bullet \mathbf{v}_1)\mathbf{p}_1\,\mathbf{v}_1.$$

(13)

Next, we present the second step to transform the second left side term in Eq. (12). Evaluating vectors $\mathbf{p}_2$ and $\mathbf{v}_1$ in $-2(\mathbf{p}_2 \bullet \mathbf{p}_3)\,\mathbf{v}_1^2$ according to Eq. (1) and Eq. (9) respectively, we obtain

$$-2(\mathbf{p}_2 \bullet \mathbf{p}_3)\,\mathbf{v}_1^2 = -2\begin{bmatrix} a^2 x_2 x_3 + a^2 y_2 y_3 + a^2 z_2 z_3 \\ b^2 x_2 x_3 + b^2 y_2 y_3 + b^2 z_2 z_3 \\ c^2 x_2 x_3 + c^2 y_2 y_3 + c^2 z_2 z_3 \end{bmatrix}.$$

(A3)

Position vector coordinates $\mathbf{p}_2$ that does not have a place in the vector in Eq. (A3) must be substituted. In this manner the variable $y_2$ and $z_2$ in the first raw, the variables in the second raw $x_2$ and $z_2$ and finally the variables in the third raw $x_2$ and $y_2$ must be changed to $x_2$, $y_2$, and $z_2$ respectively. To do this, we change variables required in each raw by using again $(x_2 - x_1)/a = (y_2 - y_1)/b = (z_2 - z_1)/c$.

$$-2(\mathbf{p}_2 \bullet \mathbf{p}_3)\mathbf{v}_1^2 = -2 \begin{bmatrix} a^2 x_2 x_3 + a^2 \left( \dfrac{b(x_2 - x_1)}{a} + y_1 \right) y_3 + a^2 \left( \dfrac{c(x_2 - x_1)}{a} + z_1 \right) z_3 \\ b^2 \left( \dfrac{a(y_2 - y_1)}{b} + x_1 \right) x_3 + b^2 y_2 y_3 + b^2 \left( \dfrac{c(y_2 - y_1)}{b} + z_1 \right) z_3 \\ c^2 \left( \dfrac{a(z_2 - z_1)}{c} + x_1 \right) x_3 + c^2 \left( \dfrac{b(z_2 - z_1)}{c} + y_1 \right) y_3 + c^2 z_2 z_3 \end{bmatrix},$$

$$= -2 \begin{bmatrix} a x_2 (a x_3 + b y_3 + c z_3) \\ b y_2 (a x_3 + b y_3 + c z_3) \\ c z_2 (a x_3 + b y_3 + c z_3) \end{bmatrix} - 2 \begin{bmatrix} a^2 (x_1 x_3 + y_1 y_3 + z_1 z_3) \\ b^2 (x_1 x_3 + y_1 y_3 + z_1 z_3) \\ c^2 (x_1 x_3 + y_1 y_3 + z_1 z_3) \end{bmatrix} + 2 \begin{bmatrix} a x_1 (a x_3 + b y_3 + c z_3) \\ b y_1 (a x_3 + b y_3 + c z_3) \\ c z_1 (a x_3 + b y_3 + c z_3) \end{bmatrix},$$

$$= -2(\mathbf{p}_3 \bullet \mathbf{v}_1)\mathbf{p}_2 \mathbf{v}_1 - 2(\mathbf{p}_1 \bullet \mathbf{p}_3)\mathbf{v}_1^2 + 2\mathbf{p}_1 \mathbf{v}_1 (\mathbf{p}_3 \bullet \mathbf{v}_1),$$

(A4)

hence,

$$-2(\mathbf{p}_2 \bullet \mathbf{p}_3)\mathbf{v}_1^2 = -2\big((\mathbf{p}_3 \bullet \mathbf{v}_1)(\mathbf{p}_2 - \mathbf{p}_1)\mathbf{v}_1 + (\mathbf{p}_1 \bullet \mathbf{p}_3)\mathbf{v}_1^2\big). \qquad (14)$$

Substituting summands of Eqs. (13) and (14) in the left side of Eq. (12) and simplifying, we obtain

$$(\mathbf{p}_2 - \mathbf{p}_1)^2 - 2(\mathbf{v}_1 \bullet (\mathbf{p}_3 - \mathbf{p}_1))(\mathbf{p}_2 - \mathbf{p}_1)\mathbf{v}_1 - (\mathbf{p}_1 \bullet (2\mathbf{p}_3 - \mathbf{p}_1))\mathbf{v}_1^2 = \left( k_1 \mathbf{v}_1 - \dfrac{n_1}{n_2}(\mathbf{p}_2 - \mathbf{p}_1) \right)^2 - (\mathbf{p}_3 \bullet \mathbf{p}_3)\mathbf{v}_1^2.$$

(A5)

This expression can be simplified and rearranged as

$$(\mathbf{p}_2 - \mathbf{p}_1)\big(\mathbf{p}_2 - \mathbf{p}_1 - 2(\mathbf{v}_1 \bullet (\mathbf{p}_3 - \mathbf{p}_1))\mathbf{v}_1\big) = \left( k_1 \mathbf{v}_1 - \dfrac{n_1}{n_2}(\mathbf{p}_2 - \mathbf{p}_1) \right)^2 - (\mathbf{p}_3 - \mathbf{p}_1) \bullet (\mathbf{p}_3 - \mathbf{p}_1)\mathbf{v}_1^2.$$

(A6)

Defining the scalar in the last term in the equation's right hand as

$$k_2 = (\mathbf{p}_3 - \mathbf{p}_1) \bullet (\mathbf{p}_3 - \mathbf{p}_1). \qquad (A7)$$

Vector quadratic equation Eq. (A6) can be expressed as

$$\mathbf{a}_1^2 - 2(\mathbf{v}_1 \bullet (\mathbf{p}_3 - \mathbf{p}_1))\mathbf{a}_1\mathbf{v}_1 = \left(k_1\mathbf{v}_1 - \frac{n_1}{n_2}\mathbf{a}_1\right)^2 - k_2\mathbf{v}_1^2,$$

$$\mathbf{a}_1^2 - 2(\mathbf{v}_1 \bullet (\mathbf{p}_3 - \mathbf{p}_1))\mathbf{v}_1\mathbf{a}_1 = k_1^2\mathbf{v}_1^2 - 2\frac{n_1}{n_2}k_1\mathbf{v}_1\mathbf{a}_1 + \frac{n_1^2}{n_2^2}\mathbf{a}_1^2 - k_2\mathbf{v}_1^2,$$
(A8)

$$n_2\mathbf{a}_1^2 - 2n_2(\mathbf{v}_1 \bullet (\mathbf{p}_3 - \mathbf{p}_1))\mathbf{v}_1\mathbf{a}_1 = n_2 k_1^2\mathbf{v}_1^2 - 2n_1 k_1\mathbf{v}_1\mathbf{a}_1 + \frac{n_1^2}{n_2}\mathbf{a}_1^2 - n_2 k_2\mathbf{v}_1^2,$$

$$\left(\frac{n_2^2 - n_1^2}{n_2}\right)\mathbf{a}_1^2 + 2\left(n_1 k_1 - n_2(\mathbf{v}_1 \bullet (\mathbf{p}_3 - \mathbf{p}_1))\right)\mathbf{v}_1\mathbf{a}_1 = n_2(k_1^2 - k_2)\mathbf{v}_1^2,$$

hence,

$$\left(\frac{n_2^2 - n_1^2}{n_2}\right)\mathbf{a}_1^2 - 2k_3\mathbf{v}_1\mathbf{a}_1 + n_2(k_2 - k_1^2)\mathbf{v}_1^2 = 0,$$
(15)

with the scalars $k_2 = (\mathbf{p}_3 - \mathbf{p}_1) \bullet (\mathbf{p}_3 - \mathbf{p}_1)$ and $k_3 = n_2(\mathbf{v}_1 \bullet (\mathbf{p}_3 - \mathbf{p}_1)) - n_1 k_1$.